\documentclass[conference]{IEEEtran}
\IEEEoverridecommandlockouts
\usepackage{cite}
\usepackage{amsmath,amssymb,amsfonts}
\usepackage{algorithmic}
\usepackage{graphicx}
\usepackage{textcomp}
\usepackage{xcolor}
\usepackage{subcaption}
\usepackage{ulem}
\usepackage{comment}

\usepackage{float} 

\def\BibTeX{{\rm B\kern-.05em{\sc i\kern-.025em b}\kern-.08em
    T\kern-.1667em\lower.7ex\hbox{E}\kern-.125emX}}
\begin{document}

\newcommand{\ar}[1]{\textcolor{blue}{#1}}
\newcommand{\as}[1]{\textcolor{olive}{ #1}}
\newcommand{\arr}[1]{\textcolor{red}{\textit{Razi: #1}}}

\title{{Adaptive Data Transport Mechanism for UAV Surveillance Missions in Lossy Environments}\\
\thanks{This material is based upon the work supported by the National Science Foundation under Grant Number 2204721 and MIT Lincoln Laboratory under Grant Number 7000612889.}
}

\author{\IEEEauthorblockN{1\textsuperscript{st} Niloufar Mehrabi}
\IEEEauthorblockA{\textit{School of computing)} \\
 \textit{Clemson University}\\
 Clemson, SC, USA\\
 nmehrab@clemson.edu}
 \and
 \IEEEauthorblockN{2\textsuperscript{nd} Sayed Pedram Haeri Boroujeni}
\IEEEauthorblockA{\textit{School of computing)} \\
 \textit{Clemson University}\\
 Clemson, SC, USA\\
 shaerib@clemson.edu}
 \and
 \IEEEauthorblockN{3\textsuperscript{rd}  Jenna Hofseth}
\IEEEauthorblockA{\textit{School of computing)} \\
 \textit{Clemson University}\\
 Clemson, SC, USA\\
 jhofset@clemson.edu}
 \and
 \IEEEauthorblockN{4\textsuperscript{th}  Abolfazl Razi}
\IEEEauthorblockA{\textit{School of computing)} \\
 \textit{Clemson University}\\
 Clemson, SC, USA\\
 arazi@clemson.edu}
 \and
 \IEEEauthorblockN{5\textsuperscript{th} James Martin}
\IEEEauthorblockA{\textit{School of computing)} \\
 \textit{Clemson University}\\
 Clemson, SC, USA\\
 jmarty@clemson.edU}
  \and
 \IEEEauthorblockN{6\textsuperscript{th} Long Cheng}
\IEEEauthorblockA{\textit{School of computing)} \\
 \textit{Clemson University}\\
 Clemson, SC, USA\\
 lcheng2@clemson.edu}
  \and
 \IEEEauthorblockN{7\textsuperscript{th} Manveen Kaur}
\IEEEauthorblockA{\textit{School of computing)} \\
 \textit{Clemson University}\\
 Los Angeles, CA, USA\\
 mkaur@calstate.edu}
  \and
 \IEEEauthorblockN{8\textsuperscript{th} Rahul Amin}
\IEEEauthorblockA{\textit{Lincoln Laboratory)} \\
 \textit{Massachusetts Institute of Technology}\\
 Lexington, MA, USA\\
 rahul.amin@ll.mit.edu}
}

\author{
    \IEEEauthorblockN{
        Niloufar Mehrabi\IEEEauthorrefmark{1},
        Sayed Pedram Haeri Boroujeni\IEEEauthorrefmark{1},
        Jenna Hofseth\IEEEauthorrefmark{1},
        Abolfazl Razi\IEEEauthorrefmark{1},
        Long Cheng\IEEEauthorrefmark{1},
        Manveen Kaur\IEEEauthorrefmark{2},\\
        James Martin\IEEEauthorrefmark{1},
        Rahul Amin\IEEEauthorrefmark{3}
    }
    \IEEEauthorblockA{
        \IEEEauthorrefmark{1}School of Computing, Clemson University, Clemson, SC, USA \\
        Email: \{nmehrab, shaerib, jhofset, arazi, jmarty, lcheng2 \}@clemson.edu
    }
    \IEEEauthorblockA{
        \IEEEauthorrefmark{2}Computer Science Department, California State University, Los Angeles, CA, USA \\
        Email: mkaur39@calstate.edu
    }
    \IEEEauthorblockA{
        \IEEEauthorrefmark{3}Lincoln Laboratory, Massachusetts Institute of Technology, Lexington, MA, USA \\
        Email: rahul.amin@ll.mit.edu
    }
}

\vspace{4 cm}

\maketitle

\begin{abstract}
Unmanned Aerial Vehicles (UAVs) play an increasingly critical role in Intelligence, Surveillance, and Reconnaissance (ISR) missions such as border patrolling and criminal detection, thanks to their ability to access remote areas and transmit real-time imagery to processing servers. However, UAVs are highly constrained by payload size, power limits, and communication bandwidth, necessitating the development of highly selective and efficient data transmission strategies. This has driven the development of various compression and optimal transmission technologies for UAVs. Nevertheless, most methods strive to preserve maximal information in transferred video frames, missing the fact that only certain parts of images/video frames might offer meaningful contributions to the ultimate mission objectives in the ISR scenarios involving moving object detection and tracking (OD/OT). This paper adopts a different perspective, and offers an alternative AI-driven scheduling policy that prioritizes selecting regions of the image that significantly contributes to the mission objective. The key idea is tiling the image into small patches and developing a deep reinforcement learning (DRL) framework that assigns higher transmission probabilities to patches that present higher overlaps with the detected object of interest, while penalizing sharp transitions over consecutive frames to promote smooth scheduling shifts. 
Although we used Yolov-8 object detection and UDP transmission protocols as a benchmark testing scenario the idea is general and applicable to different transmission protocols and OD/OT methods. To further boost the system's performance and avoid OD errors for cluttered image patches, we integrate it with inter-frame interpolations. With this method, we achieved about 45\% improvement in terms of OD accuracy for the proposed method (F1 score:98\%) compared to random selection (F1 score: 53\%) when the transmission budget is 50\% (we afford sending half of the image patches). 
Under an extremely constrained transmission budget (5\%), this gain can be as high as 87\%. The only cost for such improvement is a feedback channel from the ground server to drones, that anyway exists in most ISR scenarios for command and control purposes. Finally, we note that our method does not replace, rather it complements existing compression and video coding techniques.
\end{abstract}


\begin{figure}[htbp]
    \centering
    \includegraphics[width=0.9\columnwidth]{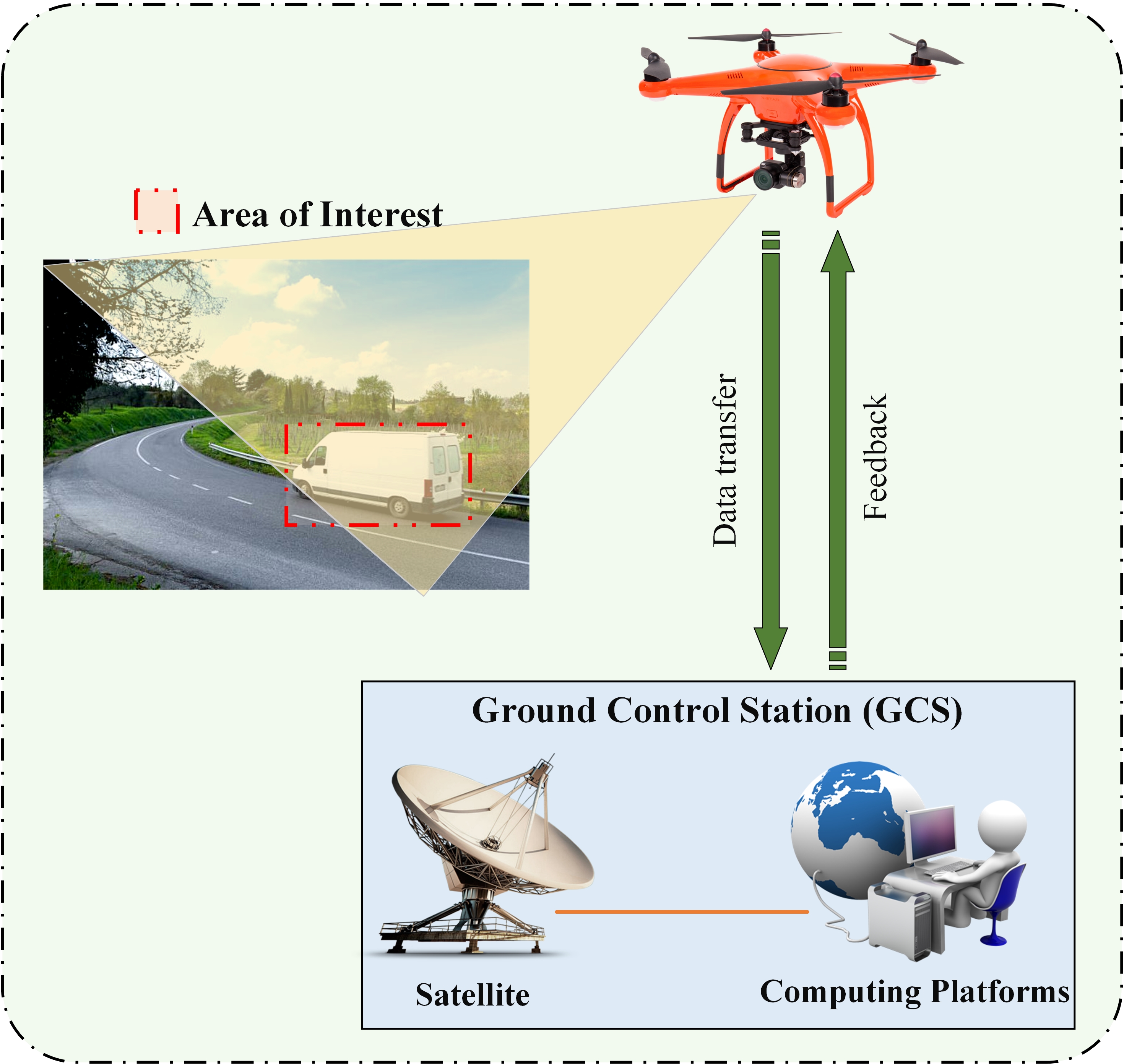}
    \caption{Overview of the UAV-based surveillance system with a feedback loop between the UAV and GCS for real-time object detection and monitoring.}
    \label{fig:Overview}
\end{figure}

\section{Introduction}



Unmanned Aerial Vehicles (UAVs) have evolved as critical components of modern cyber-physical systems, especially for providing Intelligence, Surveillance, and Reconnaissance (ISR) services. Their ability to access remote areas and provide continuous surveillance is the key motive for their widespread use in various applications, including disaster management, border security, environmental monitoring, traffic control, and search and rescue \cite{boroujeni2024comprehensive}. However, UAVs are inherently resource-constrained due to limitations in their payload and power sources. 
UAVs transform ISR data into real-time intelligence with continuous monitoring and access to dangerous areas without risking human lives. They are equipped with a variety of advanced sensors that collect crucial data on combat awareness, border security, disaster response, and many other applications, which in turn enhances decision-making and mission effectiveness.

UAVs primarily provide ISR services through video streaming, crucial for real-time decision-making. Existing video compression methods effectively reduce transmission bitrates by leveraging spatial and temporal correlations in video frames. Compression techniques such as HEVC, H.264/AVC, and VVC have been developed to minimize the loss of quality while reducing the amount of data required for video transmission \cite{ammous2023survey}.
Using various methods, most of these codecs reduce redundancy inside or between frames. For example, H.264/AVC uses block-based motion correction with transformation coding, whereas HEVC enhances these with larger block sizes and more efficient entropy coding for higher compression rates. VVC extends the capabilities even further, with up to 50\% bitrate saving compared to the HEVC, making it ideal for high-resolution video transmission over restricted bandwidth networks like those used in UAV surveillance. While these methods are effective, they might only be relevant in some ISR applications that require a focused interest in ROIs within a video frame \cite{ammous2023survey}.
For applications that require only specific parts of the image, like object detection and target tracking, selective transmission could be an effective method of transmitting only relevant parts of the image. This may bring even more impressive efficiency into compression while keeping the effectiveness of the application.

In a real-world scene, when we focus our eyes on an object, we naturally filter out irrelevant background details, which allows us to concentrate solely on the object itself. By selectively focusing on specific objects, the background does not interfere with our ability to recognize the objects, allowing us to be more adaptable to variations in the environment, such as lighting, camera angle, or background noise. However, many traditional object detection methods, including those based on Convolutional Neural Networks (CNNs) \cite{galvez2018object}, often process all parts of an image with equal emphasis. This can lead to difficulties in accurately detecting objects, as irrelevant or redundant parts of the image might introduce noise, thereby reducing the system's overall accuracy and making it more sensitive to variations in the visual environment.

Manual or rule-based selection of image regions is often inflexible and inefficient, particularly in dynamic ISR environments.  As a result, there has been a growing shift toward using Reinforcement Learning (RL) approaches, which enable more efficient and autonomous selection by continuously adapting to mission requirements in real-time.

Several studies have explored the application of RL to enhance object detection by refining the selection and processing of image regions. A notable example is the integration of RL with region selection and bounding box refinement networks, aimed at optimizing detection accuracy by refining region proposals and improving feature integration \cite{zhou2021reinforcenet}. This method focuses on optimizing detection accuracy within computational constraints. Mathe et al. \cite{mathe2016reinforcement} employ RL to optimize object detection by selectively evaluating image regions, significantly reducing computational overhead. While this approach improves detection speed, its primary focus is on computational efficiency rather than minimizing the amount of data processed. Similarly, Uzkent et al. \cite{uzkent2020efficient} use RL to optimize object detection over large images by selectively changing the spatial resolution in different image regions to reduce processing time while maintaining accuracy.
In contrast, our research emphasizes object detection with minimal data, selectively transmitting and processing only critical image segments, which is essential for resource-constrained UAV ISR missions.

In our work, we propose an adaptive communication protocol that combines the speed of User Datagram Protocol (UDP) with mechanisms to ensure that critical video data is transmitted efficiently and reliably during real-time UAV ISR missions. We leverage RL to automatically select key areas of the video frame and integrate the YOLOv8 \cite{lou2023dc} object detection algorithm to prioritize the most critical portions, ensuring they are sufficient for accurate object detection and tracking. Additionally, a feedback mechanism continuously refines the transmission strategy in response to real-time dynamic changes in network conditions and mission requirements, ensuring optimal performance.

The main contributions of this work include:
\begin{itemize}
    \item Design an RL-based intra-frame scheduling policy that assigns transmission probability to image patches based on their contribution to the mission-oriented objective. This method is integrable with optimized compression and scheduling methods to optimize resource utilization for resource-constrained UAVs.
    \item Implement an object detection that integrates inter-frame interpolation and YOLOv8 for accelerated performance.
    \item Incorporating a penalization term into the RL objective function to penalize sharp transitions to avoid abrupt policy shifts between consecutive frames.
\end{itemize}

\section{System Model}
\subsection{Network Architecture and Real-Time Data Transmission}\label{AA}

The proposed system is designed to enhance UAV-based surveillance missions by optimizing the transmission of real-time video data. A UAV captures real-time video data and transmits it frame by frame to the Ground Control Station (GCS) after some preprocessing. $F^{(n)}$ represents the $n$-th frame captured by the UAV, where each frame is divided into a grid of $K \times K$ image patches, denoted as $S_{i \times j}^{(n)}$ for $i, j = 1, 2, \ldots, K$. This fine-grained division allows for precise control over which portions of the frame are transmitted.

It uses a UDP \cite{kim2022udp} protocol to ensure low-latency communication, which is critical for real-time applications. In the proposed scheme, each patch $S_{i \times j}^{(n)}$ is transmitted as an independent packet, and the packet header includes all necessary information to reconstruct the frame $F^{(n)}$ at the GCS. The transmission probability $P_{i \times j}^{(n)}$, determined by the policy $\Pi$, dictates which patches are transmitted based on their importance. This selective transmission reduces bandwidth usage and increases transmission efficiency. Fig. \ref{fig:Overview} illustrates a UAV-based surveillance system where the UAV continuously monitors an area of interest and communicates with the GCS through a feedback loop, enabling real-time analysis and decision-making based on detected objects.

Upon receiving the packets, the GCS reassembles the frame $F^{(n)}$ from the received patches. If some patches are lost, the corresponding cells in the frame are initially replaced with black filler sections which are later replaced by interpolated estimates of the patch. This step ensures that the reconstructed image is smooth and visually coherent, enhancing the quality of the data used for further analysis.

\subsection{Deep Q-Network for Object Detection and Trajectory Prediction}

Once the GCS receives the video frames, a Deep Q-Network (DQN) is employed to analyze the content of each selected cell $\hat{S}_{i \times j}^{(n)}$. The DQN is a reinforcement learning model designed to optimize sequential decision-making, which is crucial in dynamic and time-sensitive surveillance scenarios

The system performs object recognition using a DQN in conjunction with the YOLOv8 model. Basically, YOLOv8 identifies and classifies items within grid cells, outputting bounding boxes to emphasize areas of interest. The DQN uses information to predict a trajectory, considering patterns of motion of observed objects across consecutive frames. The predictive ability allows the system to project future placements of objects, which is vital in keeping situational awareness during surveillance missions.
The DQN assigns a Q-value $Q(s_i^{(n)}, a_i^{(n)})$ for cell \(i\) in frame \(F^{(n)}\), representing the importance of the data contained within that cell. Cells that are determined to be more critical—such as those containing key objects or predicted movement paths—are assigned higher weights $w_i$. 
The Q-value is calculated for cell \(i\) in frame \(F^{(n)}\) using the Bellman equation: 

\begin{equation}
\begin{aligned}
Q(s_i^{(n)}, a_i^{(n)}) = r_i^{(n)} + \gamma Q'(s_{i+1}^{(n)}, a_{i+1}^{(n)})
\end{aligned}
\end{equation}


where \( r_i^{(n)}\) is the reward obtained by taking action \( a_i^{(n)} \) in state \( s_i^{(n)}\), \( s_{i+1}^{(n)}\) represents the next state, and \( a_{i+1}^{(n)} \) denotes possible actions in the next state. The reward function is designed to prioritize cells containing critical objects or significant motion. Also, $\gamma$ is the discount factor, which balances the importance of immediate rewards versus future rewards \cite{al2024reinforcement}.
The training of the DQN in our system is driven by a total loss function that combines two essential components: the Bellman error and a regularization term. The primary component of the loss function is the Bellman error, which is computed as the Mean Squared Error (MSE) between the predicted Q-values (\(Q(s_i^{(n)}, a_i^{(n)})\)) and the target Q-values. The target Q-value is calculated using the Bellman equation, where the immediate reward \(r_i^{(n)}\) is added to the discounted maximum expected future reward (\(\gamma Q'(s_{i+1}^{(n+1)}, a_{i+1}^{(n+1)})\)) in the subsequent frame \(F^{(n+1)}\). This term encourages the DQN to approximate the optimal action-value function by minimizing the difference between the current Q-values and the target Q-values. In addition to the Bellman error, the loss function includes a regularization term that penalizes significant changes in the action probabilities over consecutive time steps. This regularization is critical for maintaining stability in the learning process, as it discourages abrupt shifts in the policy that could lead to erratic behavior. The strength of this regularization is controlled by the parameter \(\lambda\), which balances the trade-off between fitting the Q-values accurately and ensuring smooth transitions in the learned policy. The total loss function is therefore expressed as:

\begin{equation}
\begin{aligned}
\text{Total Loss} &= \frac{1}{N} \sum_{n=1}^{N} \sum_{i=1}^{M} \Bigl(Q(s_i^{(n)}, a_i^{(n)}) \\ & -
(r_i^{(n)} + \gamma Q'(s_{i+1}^{(n+1)}, a_{i+1}^{(n+1)}) ) \Bigl) ^2 \\  
& + \lambda \sum_{n=2}^{N} \sum_{j=1}^{|A|} \left( P_j^{(n)} - P_j^{(n-1)} \right)^2
\end{aligned}
\end{equation}



Here, N is the total number of frames in the sequence, and M represents the total number of grid cells in each frame. Also, the first term is the MSE between the predicted and target Q-values, while the second term represents the change in probability for action \(j\) between frame \(F^{(n)}\) and frame \(F^{(n\_1)}\). By summing this across all actions and frames, the regularization term enforces smoother transitions. By minimizing this loss, the DQN learns to predict future states while ensuring stability and robustness, vital for the dynamic and time-sensitive nature of UAV-based surveillance missions.

Once the Q-values are calculated, the system assigns weights \( w_i \) to each selected cell $\hat{S}_{i \times j}^{(n)}$ in the grid. The weight for cell $\hat{S}_{i \times j}^{(n)}$ can be represented as:

\begin{equation}
\begin{aligned}
w_i = \frac{Q(s_i, a_i)}{\sum_{j=1}^{K^2} Q(s_j, a_j)}
\end{aligned}
\end{equation}

This normalization ensures that the weights are proportional to the relative importance of each cell, with higher weights indicating cells that are more critical for transmission.

\subsection{Feedback Loop and Adaptive Transmission}

The feedback loop is crucial for enhancing the data transmission process. After the DQN assigns weights $w_i$ to the cells, this information is fed back into the system to update the transmission strategy. The UAV uses feedback information to determine which cells are most important and should be sent to the GCS in the next transmission cycle.
Based on the weights \( w_i \), the feedback loop guides the UAV in prioritizing specific cells for the next transmission cycle. The transmission probability \( P_i \) for each cell can be defined as:

\begin{equation}
\begin{aligned}
P_i = \frac{w_i}{\sum_{j=1}^{K^2} w_j}
\end{aligned}
\end{equation}

This probabilistic approach ensures that the transmission strategy is adaptive and focused on the most important data. 
As the DQN learns, the feedback loop is continuously refined. In order to improve future performance, the system monitors the results of its transmission strategies, such as the success rate of data delivery and the consistency of object detection.
As a result of this adaptive approach, the system maintains high levels of efficiency and accuracy in real-time surveillance, making it especially useful for ISR missions. The combination of dynamic data prioritization, adaptive transmission, and real-time feedback ensures that the most critical information is always available to decision-makers, even in challenging environments.

\section{Experiment Results}

\subsection{Dataset}
\begin{figure}[H]
    \centering
    \includegraphics[width=1\columnwidth]{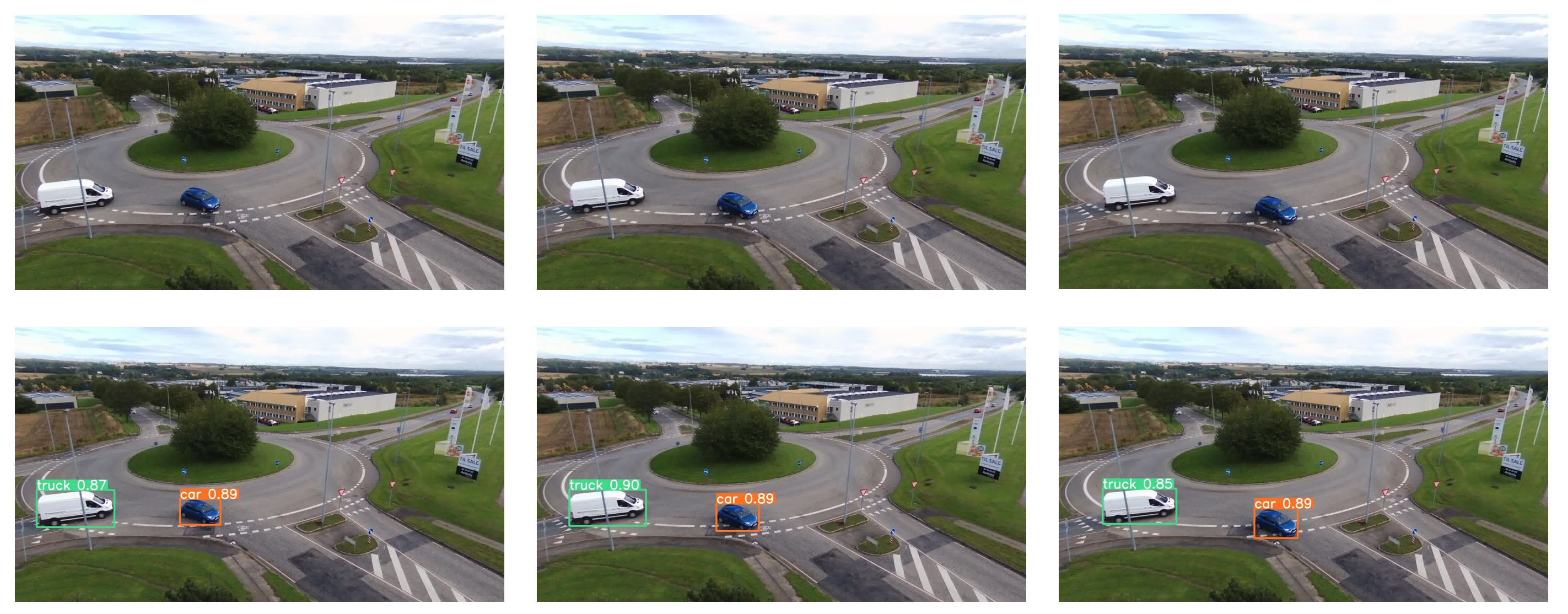}
    \caption{Sample frames and labeled with object annotations from the AU-AIR dataset.}
    \label{fig:datasample}
\end{figure}

The experiments were conducted using the AU-AIR dataset, which is a comprehensive dataset specifically designed for aerial object detection tasks. The AU-AIR dataset is made for research in computer vision and autonomous systems, focusing on UAVs \cite{bozcan2020air}. The AU-AIR dataset is built on high-resolution video sequences acquired from a UAV in various real-world environments. The dataset contains more than 32,000 annotated frames containing a variety of objects, like vehicles, pedestrians, cyclists, and fixed object classes, such as traffic signs. Fig. \ref{fig:datasample} illustrates a sample frame extracted from the AU-AIR dataset, showcasing the typical objects and scenes encountered during the UAV's surveillance missions \cite{boroujeni2024ic}.

\subsection{Training Phase}
\begin{figure}[H]
    \centering
    \includegraphics[width=\columnwidth]{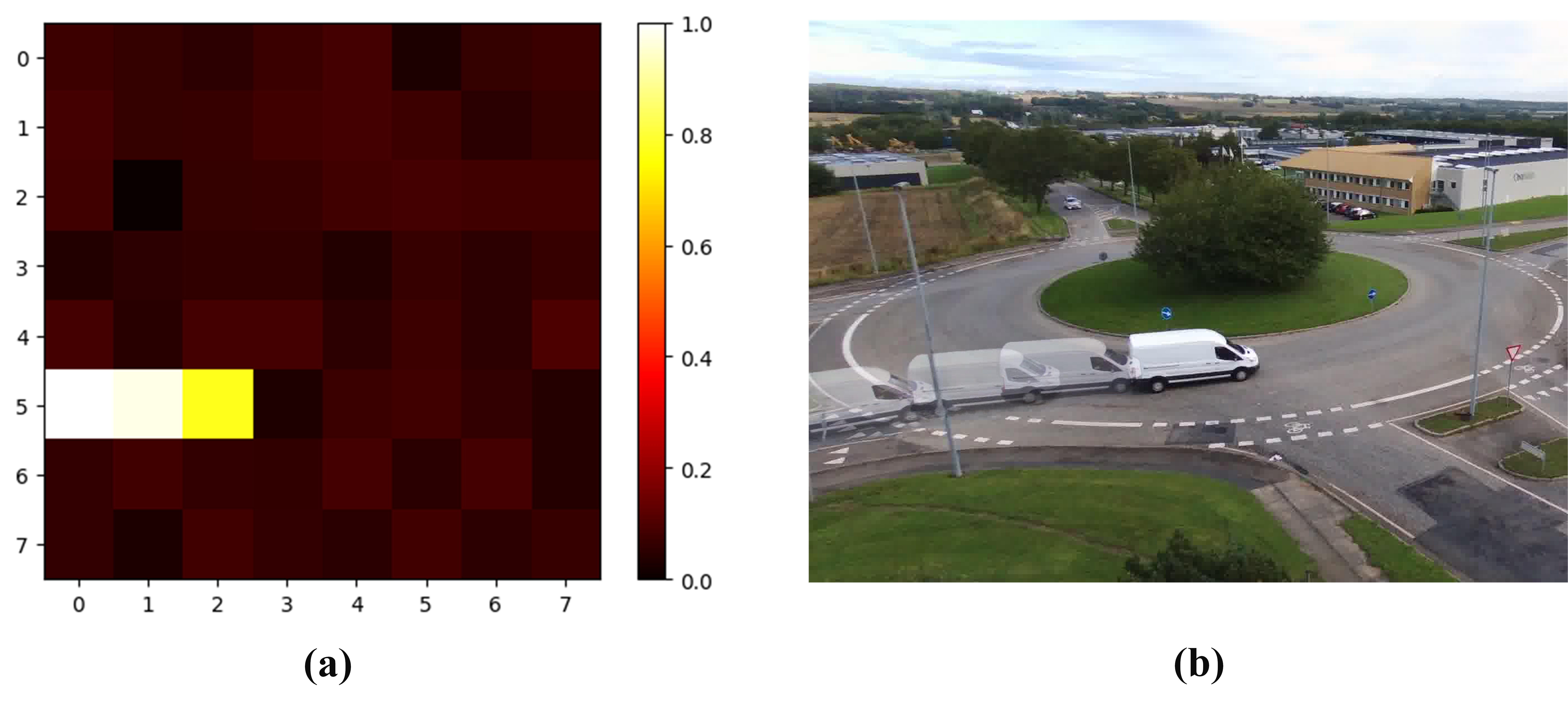}
    \caption{Visualization of the DQN training results: (a) A heatmap representing the varying importance of different regions within the frame, where lighter cells indicate higher importance or the likelihood of object presence as predicted by the DQN. (b) The truck's path, visualized by blending sequential frames, demonstrates the DQN's ability to accurately predict the movement trajectory based on past observation}
    \label{fig:DQN}
\end{figure}

Initially, the UAV sends the first four complete frames to the server using UDP. Those frames are used to train a DQN combined with a YOLOv8, which is responsible for detecting and recognizing target object in the frames. Our primary focus is on Single Object Tracking (SOT). The DQN is trained by received frames to learn the importance of different regions in the frames, generating a probability distribution across the grid cells. The results of this training process are visualized through a heatmap that illustrates the varying importance of different regions within the frame, as shown in Fig. (\ref{fig:DQN}.a). Lighter cells indicate the presence of an object or the likelihood of objects moving through these paths, as predicted by the DQN. Fig. (\ref{fig:DQN}.b) shows the truck's path, created by blending some sequential frames, which visually represent its movement. This path is then accurately predicted by the DQN, highlighting its effectiveness in tracking and forecasting the truck’s movements based on previous frames.

\begin{figure*}[h]
    \centering
    \includegraphics[width=0.95\textwidth]{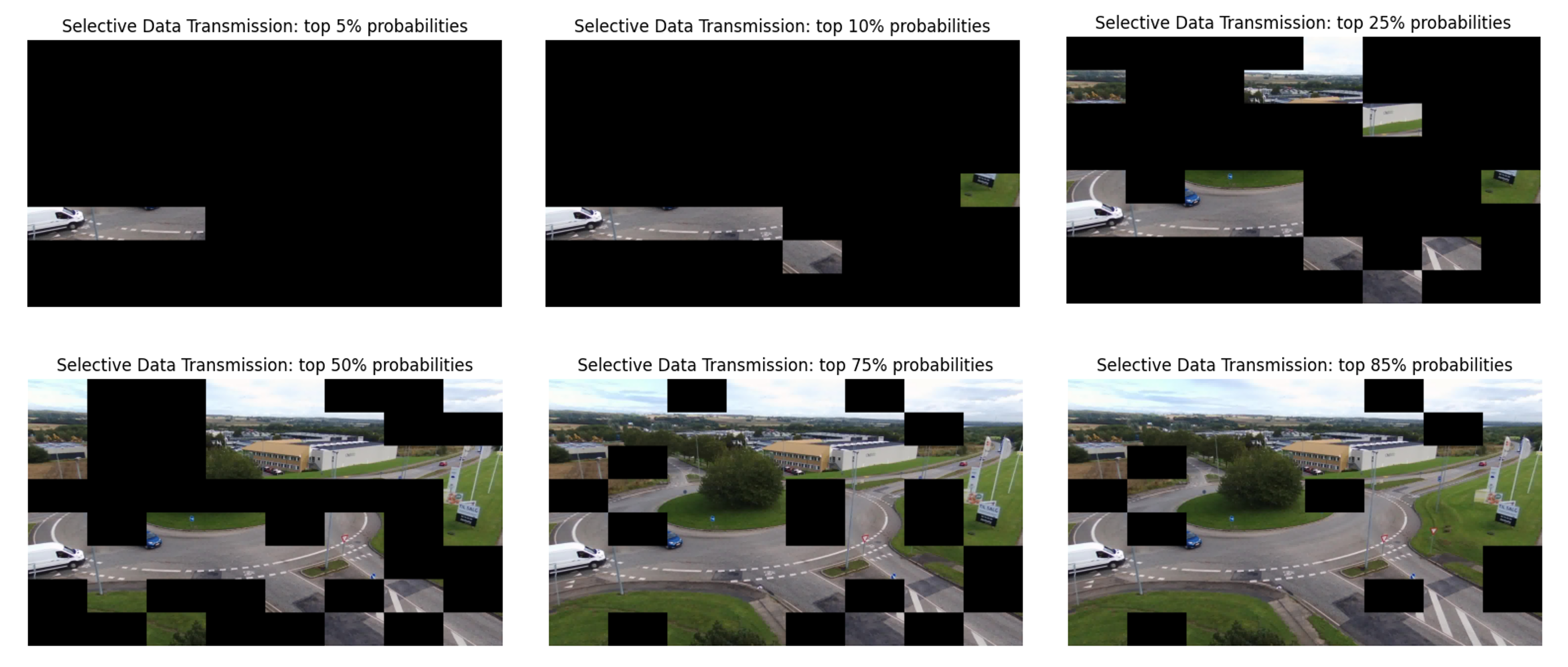}
    \caption{Selected cells for transmission based on the top probabilities using different masks.  }
    \label{fig:selectivedata}
\end{figure*}

The binary mask generated based on obtained probabilities from the DQN highlights the most critical cells in the frame, ensuring that only the most important regions are selected for transmission, thus optimizing the use of available bandwidth. This mask is then sent back to the UAV as feedback. The feedback mechanism allows the UAV to focus on transmitting only the most critical parts of the frame in subsequent transmissions based on the obtained probabilities.
In every consecutive frame, the UAV divides the image into $K\times K$ (K=8) cells and further uses a mask to decide how many cells will be transmitted to the server. The selection is based on varying percentages (5\%, 10\%, 25\%, 50\%, 75\%, and 85\%) of the total cells, s, chosen according to the highest probabilities, reflecting different levels of data prioritization and bandwidth usage. The selected cells $\hat{S}_{i \times j}^{(n)}$ for a single frame, aimed at detecting the truck as the target object, are illustrated in Fig. \ref{fig:selectivedata}. 


\begin{figure}[h]
    \centering
    \includegraphics[width=\columnwidth]{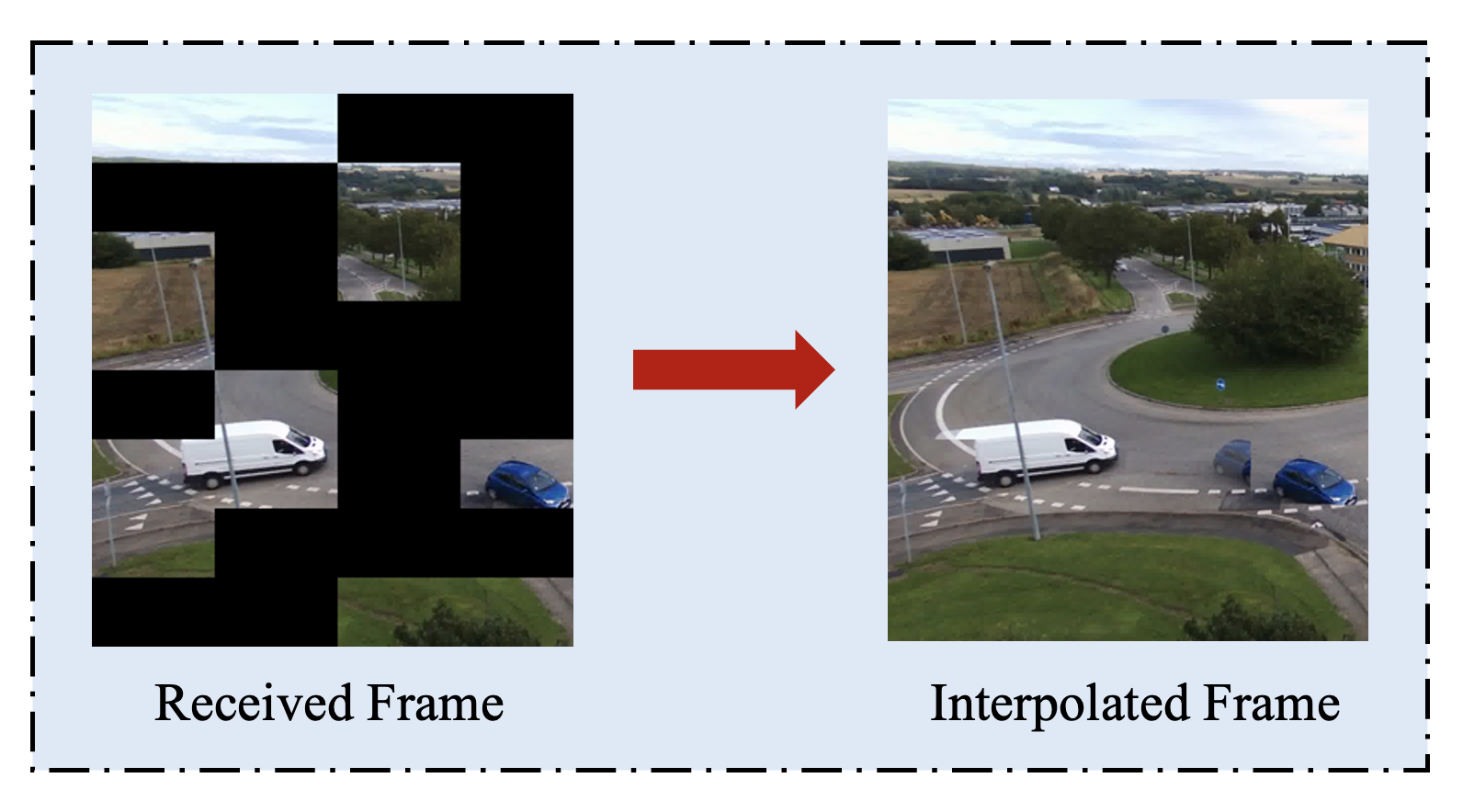}
    \caption{Illustration of the frame reconstruction process: the received frame, containing only selectively transmitted cells, is shown on the left. The right image demonstrates the result after applying interpolation to the missing areas, effectively reconstructing a more coherent and visually complete frame.}
    \label{fig:interpolated}
\end{figure}

Upon receiving the selected cells, the server reconstructs the image with placeholders for missing cells, then applies interpolation to improve visual quality, as shown in Fig. \ref{fig:interpolated}. The black placeholders indicate the missing cells, while the smooth transitions highlight the effectiveness of the interpolation method.

The DQN and the mask are continually updated in this process, since new frames are added to the processing queue; hence, the system allows a constant updating of which parts of the image become the most important. The feedback loop ensures that the UAV can dynamically adapt its transmission strategy based on the updated information.
 
\subsection{Evaluation and Performance Analysis}

\begin{figure}[htbp]
    \centering
    \includegraphics[width=\columnwidth]{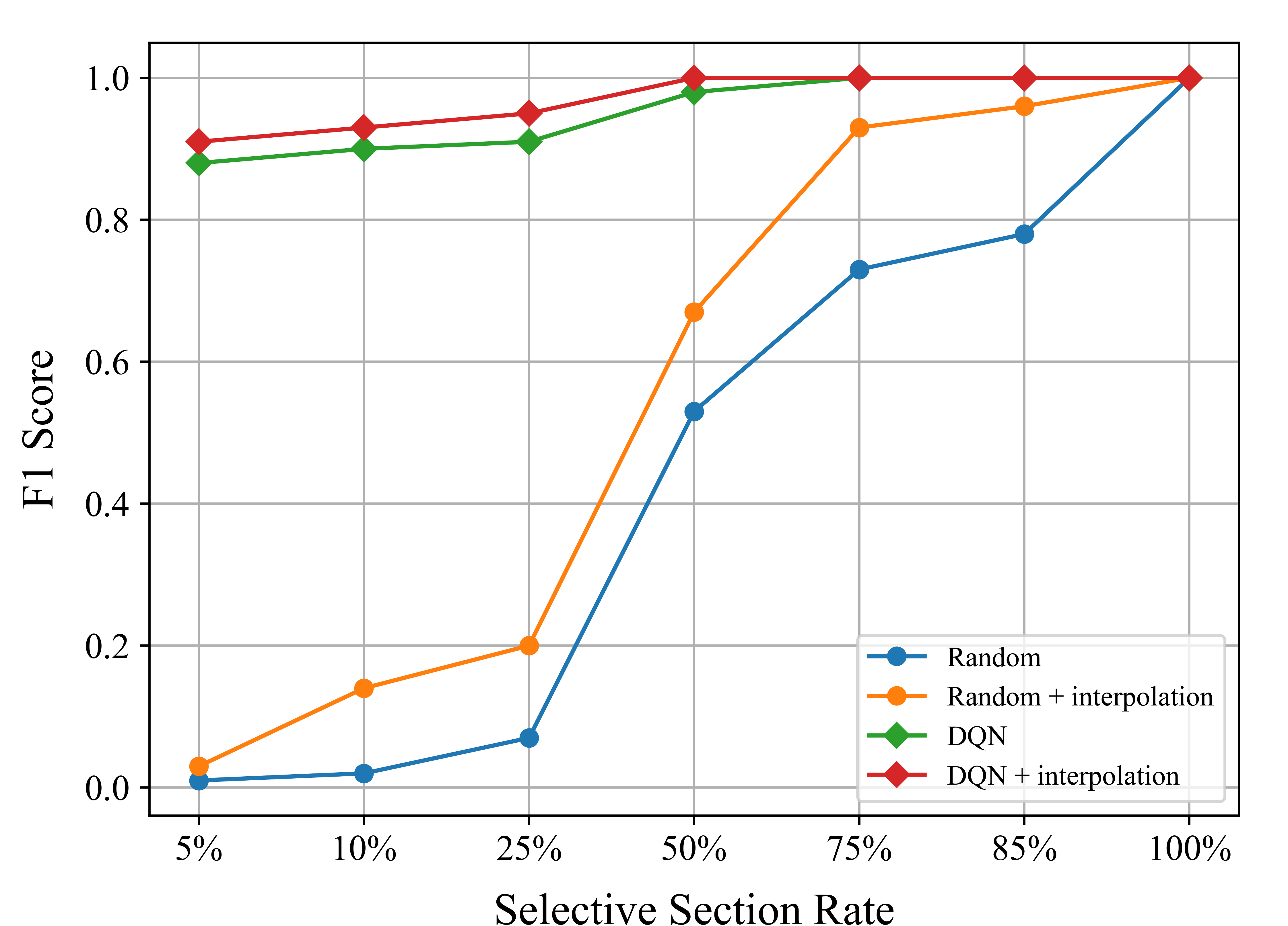}
    \caption{Comparison of F1 scores across different transmission rates.}
    \label{fig:F1}
\end{figure}

\begin{figure}[htbp]
    \centering
    \includegraphics[width=\columnwidth]{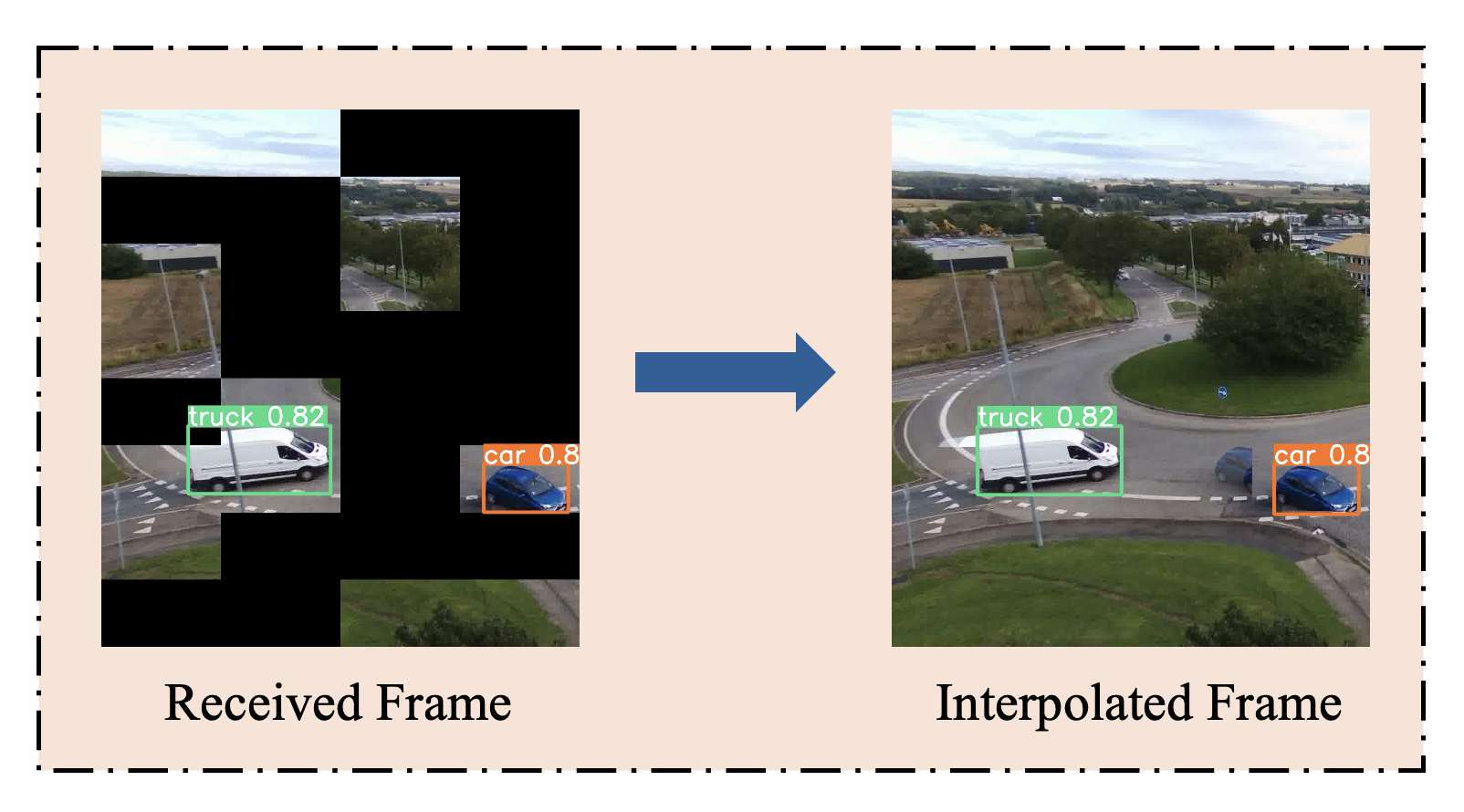}
    \caption{Object detection results on both the received frame and the interpolated frame.}
    \label{fig:od}
\end{figure}

\begin{figure}[htbp]
    \centering
    \includegraphics[width=\columnwidth]{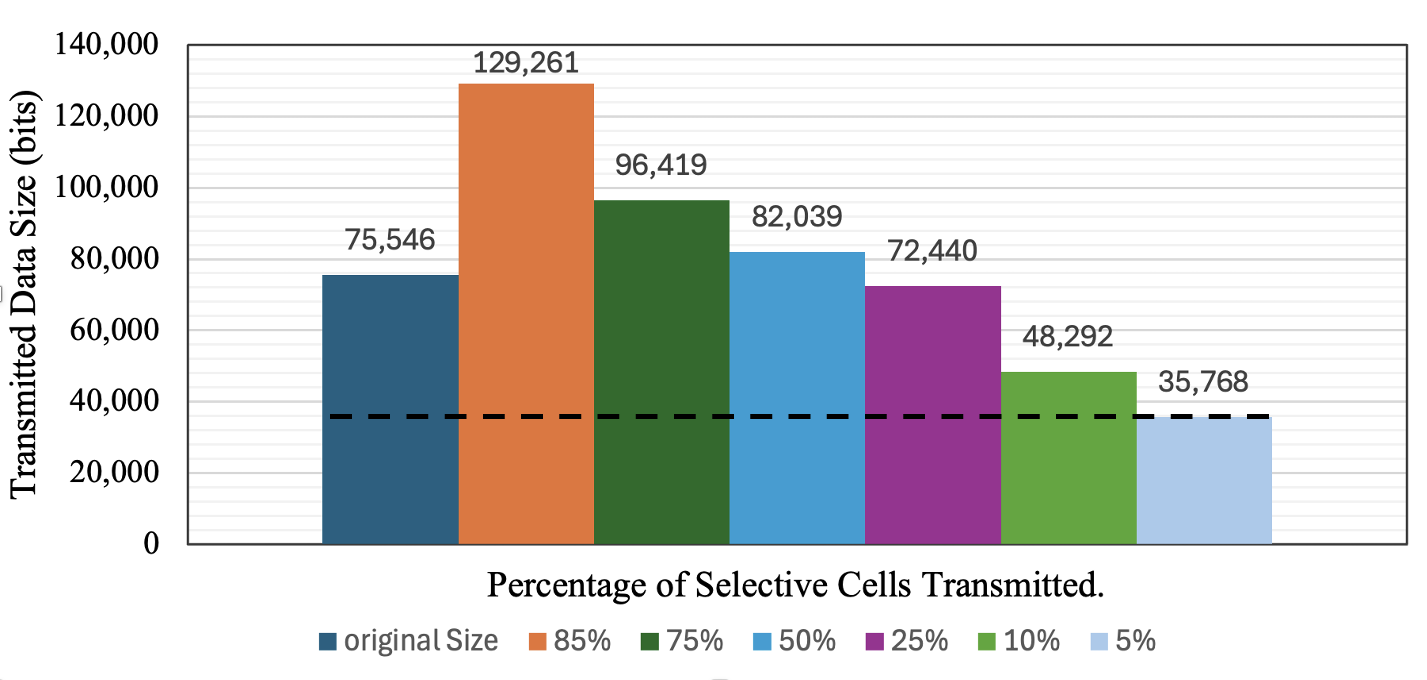}
    \caption{Comparison of Transmitted Data Size (bits) between sending the entire image and independently sending split cells at varying percentages.}
    \label{fig:bitusage}
\end{figure}

One of the key advantages of the proposed system is its ability to reliably detect and track objects with minimal data transmission. In this regard, we tested the system's object detection performance for various selective section rates, which were between 5\% and 85\% of the total number of grid cells based on the highest probability scores. Object detection performance was quantified at each rate using the F1 score and Precision. 
Fig. \ref{fig:F1} clearly demonstrates how the proposed DQN-based strategy, especially when combined with interpolation, maintains strong accuracy even when transmitting only selective sections of the image, as low as 5\% of the grid cells. The corresponding details are also included in Table \ref{tab:performance_metrics} for completeness. 
The proposed DQN-based selective transmission technique is effective since it focuses on the most important regions for object recognition, hence carrying out detection and tracking in a very efficient manner with less data. Thus, this facility for detecting and tracking objects with such minimal data reduces the bandwidth requirement and enhances the efficacy and suitability of the proposed method for real-time UAV surveillance in bandwidth-constrained environments, as demonstrated in Fig. \ref{fig:od}. Furthermore, Fig. \ref{fig:bitusage} illustrates the transmitted data size at different selective section rates, demonstrating the efficiency of the proposed method in maintaining low bandwidth requirements while still enabling effective object detection.

\begin{table}[htbp]
\caption{Performance Metrics for Various Selective Section Rate}
\label{tab:performance_metrics}
\renewcommand{\arraystretch}{2}  
\begin{center}
\resizebox{\columnwidth}{!}{%
\begin{tabular}{|c|c|c|c|c|c|}
\hline
\textbf{Selective} & \textbf{Performance} &\multicolumn{4}{|c|}{\textbf{Methods}} \\
\cline{3-6} 
\textbf{Section Rate} & \textbf{Metrics} & \textbf{\textit{Random}} & \textbf{\textit{Random+Interpolation}} & \textbf{\textit{DQN}} & \textbf{\textit{DQN+Interpolation}} \\
\hline

5\% & F1         & 1\%  & 3\%  & 88\% & 91\% \\
    & Precision  & 1\%  & 3\%  & 88\% & 91\% \\
\hline
10\% & F1         & 2\%  & 14\% & 90\% & 93\% \\
     & Precision  & 2\%  & 14\% & 90\% & 93\% \\
\hline

25\% & F1         & 7\%  & 20\% & 92\% & 95\% \\

     & Precision  & 7\%  & 20\% & 92\% & 95\% \\
\hline
50\% & F1         & 53\% & 67\% & 98\% & 100\% \\
     & Precision  & 53\% & 67\% & 98\% & 100\% \\
\hline

75\% & F1         & 73\% & 93\% & 100\% & 100\% \\

     & Precision  & 73\% & 93\% & 100\% & 100\% \\
\hline
85\% & F1         & 78\% & 96\% & 100\% & 100\% \\
     & Precision  & 78\% & 96\% & 100\% & 100\% \\
\hline

100\% & F1        & 100\% & 100\% & 100\% & 100\% \\

      & Precision & 100\% & 100\% & 100\% & 100\% \\
\hline
\end{tabular}%
}
\end{center}
\end{table}

\section{Conclusion}

In this work, we proposed a DQN-based selective transmission approach to improve object detection and tracking in real-time UAV surveillance missions. The current practice is developing optimal video coding, compression, and transmission strategies for resource-constrained UAVs to minimize the required bitrate for video streaming under dynamic conditions while remaining loyal to transmitting the entire video frames. Here, we take a different approach and develop an intra-frame scheduling policy, which prioritizes the transmission of image parts that make meaningful contributions to the ultimate mission objective. 
The quality of the received frames could be further improved by incorporating interpolation that made it possible to do effective object recognition with least amount of data. The results obtained from experiments illustrate that the proposed approach 
yields dependable performance in constrained communication environments. The gain in the object detection and tracking accuracy is about 45\% with respect to random selection when the transmission budget is 50\% (F1 score). This gain can be as high as 90\% for extremely constrained transmission budget (5\%). The cost for this gain is the existence of a lightweight feedback channel from the ground server to UAVs. The processing delay is negligible and our method supports real-time video streaming with 30 FPS (frame per second).  These results reveal the efficacy of the proposed system in enhancing aerial surveillance systems. 
More has to be done to embed this method into modern video coding and compression techniques. This research can be further enhanced through the integration of the proposed method with modern video coding and techniques in compression for better data transmission efficiency. Moreover, the extension of Multi-Object Tracking (MOT) would further strengthen the capability of this system in terms of the complexity of surveillance scenarios, making it more apt for multi-target handling.



\bibliographystyle{ieeetr}
\bibliography{ref} 

\end{document}